\documentclass[12pt,letterpaper]{article}

\usepackage{amsmath,amssymb,calc}
\usepackage{graphicx}
\usepackage{color}

\newcommand\be{\begin{equation}}
\newcommand\ee{\end{equation}}
\newcommand{\bea}{\begin{eqnarray}}
\newcommand{\eea}{\end{eqnarray}}

\newcommand{\nn}{\nonumber}
\newcommand{\pd}{\partial}

\def\id{\protect{{1 \kern-.28em {\rm l}}}}

\def\m{\mu}

\def\1{^{(1)}}
\def\0{^{(0)}}
\def\2{^{(2)}}

\def\id{\protect{{1 \kern-.28em {\rm l}}}}

\let\non\nonumber

\begin{document}

\begin{titlepage}
\begin{center}
\hfill \\

\vskip 2cm
{\Large \bf On the stability of D7-${\bf \overline{{\rm {\bf D}}7}}$ probes in \vspace{0.2cm}\\ near-conformal backgrounds}

\vskip 1.5 cm
{\bf  Lilia Anguelova${}^a$\footnote{languelova@perimeterinstitute.ca}, Peter Suranyi${}^b$\footnote{peter.suranyi@gmail.com} and L.C.R. Wijewardhana${}^b$\footnote{rohana.wijewardhana@gmail.com}\\
\vskip 0.5cm  {\it ${}^a$ Perimeter Institute for Theoretical Physics, Waterloo, ON N2L 2Y5, Canada\\ ${}^b$ Dept. $\!$of Physics, University of Cincinnati,
Cincinnati, OH 45221, USA}\non\\}

\end{center}
\vskip 2 cm
\begin{abstract}
We investigate the perturbative stability of a nonsupersymmetric D7-$\overline{{\rm D}7}$ brane embedding in a particular class of type IIB backgrounds. These backgrounds are the gravitational duals of certain strongly-coupled gauge theories, that exhibit a nearly conformal regime (known as a walking regime). Previous studies in the literature have led to conflicting results as to whether the spectrum of fluctuations around the flavor D7-$\overline{{\rm D}7}$ embedding has a tachyonic mode or not. Here we reconsider the problem with a new analytical method and recover the previously obtained numerical results. We also point out that the earlier treatments relied on a coordinate system, in which it was not possible to take into account fluctuations of the point of confluence of the D7-$\overline{{\rm D}7}$ branes. Using an improved coordinate system, we confirm the presence of a normalizable tachyonic mode in this model, in agreement with the numerical calculations. Finally, we comment on the possibility of turning on worldvolume flux to stabilize the probe branes configuration.

\end{abstract}
\end{titlepage}

\tableofcontents

\section{Introduction}

Theories that are nearly conformal for a large range of energy scales, such as walking gauge theories \cite{WalkTech}, could exhibit non trivial dynamics, but are notoriously hard to analyze due to their strong coupling. Gauge/gravity duality \cite{GGD,GGDrev,GGDmore_gen}, which maps quantum dynamics of strongly coupled gauge theories to classical solutions of gravitational theories in higher dimensions, is a promising technique to analyze such models. Using this method, we studied recently a model of a walking gauge theory, that may be relevant for phenomenological applications related to the LHC. More specifically, in \cite{LA} one of us introduced U-shaped stacks of D7-$\overline{{\rm D}7}$ branes to incorporate chiral symmetry breaking, a la Sakai-Sugimoto \cite{SS}, in the gravity background of \cite{piai}. Subsequently, we investigated the spectrum of vector \cite{ASW1} and scalar mesons \cite{ASW2} in this model. The scalar boson spectrum was obtained from studying the fluctuations of the D7-$\overline{{\rm D}7}$ branes. Using an approximate linearized version of the relevant equations of motion, we obtained a spectrum of fluctuations with all eigenvalues $E=m^2$ positive, corresponding to a stable embedding of the branes. It turns out, however, that this consideration did not capture the full picture. 

The position of the D7-$\overline{{\rm D}7}$ embedding in the transverse space can be described by two angles $\theta$ and $\varphi$, which are functions of a radial variable $z$ running over both the D7 and $\overline{{\rm D}7}$ branes. The equations of motion for the scalar mesons, arising from fluctuations around that embedding, are second-order coupled differential equations for the deviations $\delta\theta$ and $\delta\varphi$ from the classical solution. Those equations contain first derivatives. However, they can easily be brought to a Schr\"odinger form by simple transformations. In a recent paper, the spectra of these Schr\"odinger equations were analyzed with numerical integration techniques \cite{CLtV}. That work found a state with negative mass squared, i.e. a tachyon, in the $\delta \varphi$ spectrum when Neumann boundary conditions were imposed at $z=0$.  The tachyon appears if the cutoff in the variable $z$, denoted by $z_\Lambda$ \footnote{A  cutoff, such that $-z_\Lambda\leq z\leq z_\Lambda$, must be used in the model to have a discrete spectrum.}, exceeds a critical value $z_c\simeq 1.5$.  Note that a large value of $z_\Lambda$ is required to have slowly running coupling constant. In this note, we will explain how the results of \cite{ASW2} and \cite{CLtV} are compatible with each other. Furthermore, we will introduce an elegant new coordinate system that allows one to study the fluctuation of the tip of the embedding, unlike the coordinates used in both \cite{ASW2,CLtV}. In the new system, some coordinates are transverse and others longitudinal {\it everywhere} along the D7-$\overline{{\rm D}7}$ embedding. Note that any other coordinate system would lead to the same  physical results, as long as the fluctuating field has a nonzero transverse component {\it everywhere} along the probe branes.\footnote{The field $\varphi$, used in \cite{ASW2,CLtV} to describe transverse fluctuations, does not satisfy this condition in a neighborhood of the tip of the flavor branes embedding. We are grateful to T. ter Veldhuis for useful correspondence, that spurred our improved understanding of this issue.} Using the new coordinates, we will verify the presence of the tachyonic mode in this model. As in other cases in the literature (see for ex. \cite{MW,BJLL}), it may be possible to remove such a mode by turning on background flux for the worldvolume gauge field. We will leave this issue for a future study.  

First, in the next section we recapitulate the results of \cite{ASW2} concerning the spectrum of $\delta\varphi$ scalar bosons. Then, in Section 3 we introduce an analytical method, relying on the study of zero mass states and a relationship they have to tachyons, to find exact conditions for the existence of negative energy states (tachyons) in the Schr\"odinger equation. Our results are in complete agreement with the numerical results of \cite{CLtV} for the $\delta \varphi$ sector containing the tachyon, namely the modes with Neumann boundary condition at $z=0$. In addition, we find a transcendental equation, whose solution provides the value of $z_c$. This is also in agreement with the numerical result of \cite{CLtV}.\footnote{We should note that, while our considerations use explicitly the metric of the walking region, the numerical investigation of \cite{CLtV} utilizes the full metric. This, though, is not relevant for the discussion of the instability.} Despite that, we point out that the tachyonic mode of the Schr\"odinger equation, when transformed back to a solution of the original equation, becomes a singular mode. This would seem to imply that it is ruled out by the regularity condition needed for the perturbative spectrum. However, we will show in a later section that the singularity of the tachyonic mode was due to an inappropriate choice of coordinate system.  

In Section 4 we reconsider the $\delta \theta$ spectrum with the new method developed here, that is based on the investigation of zero mass modes of the equations of motion. We find instabilities in this sector both for Neumann and Dirichlet boundary conditions at $z=0$. However, unlike the instability in the $\delta\varphi$ mode, these instabilities satisfy the regularity condition. Nevertheless, they are not expected to be present in phenomenologically relevant models. The reason is that, in the parameter range required to have a slowly running gauge coupling, these instabilities occur when the cutoff $z_{\Lambda}$ is chosen to be extremely large. Such a choice of $z_{\Lambda}$, however, is incompatible with the phenomenological constraints in this kind of model \cite{ASW1}.\footnote{Note that these constraints were derived from the vector meson spectrum, which does not contain a tachyon. One may expect the inclusion of background flux for the worldvolume gauge field, needed to remove the scalar spectrum tachyon, to preserve the previous conclusions regarding the phenomenological constraints. However, it should be kept in mind that this issue merits a separate investigation.} 

In Section 5 we argue that the field $\varphi$ used so far, here as well as in the previous literature, does not capture the full scalar spectrum. The reason is that it does not describe a transverse direction everywhere along the probe D7-$\overline{{\rm D}7}$ embedding. Hence, it may miss information about the scalar meson spectrum, which is supposed to arise from the transverse fluctuations. We then introduce appropriate new coordinates and show that there is a physical tachyonic mode, signaling pertubative instability.\footnote{In fact, up to the Jacobian of the relevant transformation, that mode agrees with the result of \cite{CLtV}.} We point out that the model could be stabilized by turning on worldvolume flux, in the vein of \cite{MW}, but we leave this interesting question for the future. 

In Appendix A we discuss in more detail the singular solution in the $\delta \varphi$ sector. The singularity occurs because the transformation between the Schr\"odinger equation and the original equation of motion is singular at $z=0$. This leads to $\delta \varphi$ having a pole at $z=0$. We recall two separate arguments for requiring the physical modes to be regular.

In Appendix B we prove a set of theorems, that show the relationship between zero energy solutions and negative energy ones in a Schr\"odinger equation with a cutoff. These are needed to obtain the results of Sections 3 and 4. We should note that the new method for studying questions of (in)stability, that arises from this appendix, is a rather powerful tool for investigating nonsupersymmetric brane embeddings in nontrivial backgrounds. 

\section{Set-up and first look at stability issue}
\setcounter{equation}{0}

The system of interest for us arises from considering probe D7 branes embedded into the background of \cite{piai}. The latter background is an $N=1$ supersymmetric solution of type IIB, sourced by a certain stack of D5 branes. The metric in that background depends on two parameters $\alpha$ and $c$. The solution of \cite{piai} is given as an expansion in the parameter $\frac{1}{c} <\!\! <1$ and, to leading order, has the form:
\bea
ds^2 \!&=& \!A \left[ dx_{1,3}^2 + \frac{cP_1' (\rho)}{8} \left( 4 d\rho^2 + (\omega_3 + \tilde{\omega}_3)^2 \right) \right. \nn \\
&+& \!\left. \frac{c\,P_1 (\rho)}{4} \left( \frac{1}{\coth(2\rho)} d\Omega_2^2 + \coth(2\rho) d\tilde{\Omega}_2^2 + \frac{2}{\sinh ( 2\rho)} (\omega_1 \tilde{\omega}_1 - \omega_2 \tilde{\omega}_2) \right) \right]\!,
\eea
where
\be \label{AP1}
A=\left( \frac{3}{c^3 \sin^3 \alpha} \right)^{1/4} , \quad  P_1' (\rho) =\frac{\pd P_1 (\rho)}{\pd \rho} \,\, , \quad P_1 (\rho) = \left( \cos^3 \alpha + \sin^3 \alpha \left( \sinh (4 \rho) - 4 \rho \right) \right)^{1/3} \, ,
\ee
\bea \label{omegatilde}
\tilde{\omega}_1 &=& \cos \psi d\tilde{\theta} + \sin \psi \sin \tilde{\theta} d \tilde{\varphi} \,\, , \hspace*{2cm} \omega_1 = d \theta \,\, , \nn \\
\tilde{\omega}_2 &=& - \sin \psi d\tilde{\theta} + \cos \psi \sin \tilde{\theta} d \tilde{\varphi} \,\, , \hspace*{1.6cm} \omega_2 = \sin \theta  d \varphi \,\, , \nn \\
\tilde{\omega}_3 &=& d \psi + \cos \tilde{\theta} d \tilde{\varphi} \,\, , \hspace*{3.7cm} \omega_3 = \cos \theta d \varphi
\eea
and
\be
d\tilde{\Omega}_2^2 = \tilde{\omega}_1^2 + \tilde{\omega}_2^2 \,\, , \hspace*{2cm} d\Omega_2^2 = \omega_1^2 + \omega_2^2 = d \theta^2 + \sin^2 \theta d \varphi^2 \,\, .
\ee
There is also nontrivial RR 3-form flux, which is not relevant for the following.\footnote{Note also that, to leading order in $1/c$, the string dilaton is constant in this solution.} 

This background is the gravitational dual to a walking gauge theory. In the walking regime, the above metric simplifies to \cite{ASW2,ASW3}:
\be \label{leadMetric}
ds^2_{{\rm walk}} = A \left[ \eta_{\mu \nu} dx^{\mu} dx^{\nu} + \frac{c}{12} \beta e^{4 \rho} \left( 4 d\rho^2 + \left( \omega_3 + \tilde{\omega}_3 \right)^2 \right) + \frac{c}{4} \left( d \Omega_2^2 + d \tilde{\Omega}_2^2 \right) \right] ,
\ee 
where we have introduced the notation
\be
\beta \equiv \sin^3 \alpha <\!\!< 1 \, .
\ee
A realistic model of dynamical electroweak symmetry breaking would need to encode, at energies (radial distances) above the walking region, a larger symmetry group known as extended technicolor \cite{ExtTech}. Modifying the above solution, to take this into account, is still an open problem, although there has been recent progress in that direction \cite{Cascade}. So, for practical reasons, we have introduced a cut-off $\rho_{\Lambda}$ \cite{ASW1}, which is the physical scale corresponding to the symmetry breaking of that larger gauge group to the smaller group of the walking regime. Our model, thus, uses the background (\ref{leadMetric}) with $\rho \le \rho_{\Lambda}$.

To introduce flavor degrees of freedom and study dynamical chiral symmetry breaking in the above set-up, one can show that there is a U-shaped embedding, similar to \cite{SS,KS}, of D7 $-$ anti-D7 probe branes in (\ref{leadMetric}); see \cite{LA}. Let us denote the radial position of the tip of that embedding by $\rho_0$.\footnote{Note that the shape of the embedding around the tip is {\it smooth}.} In \cite{ASW2} we introduced for convenience a new radial variable by:
\be
z = \pm \sqrt{e^{4(\rho-\rho_0)} - 1} \, .
\ee
The conceptual difference between $\rho$ and $z$ is that $\rho$ is a spacetime radial coordinate, whereas $z$ runs only over the worldvolume of the D7-$\overline{{\rm D}7}$ embedding. The shape of the embedding in the two transverse directions is described by the fields $\theta (z)$ and $\varphi (z)$, which parametrize the position of a worldvolume point in the transverse $(\theta,\varphi)$ two-sphere. The scalar mesons in this model arise from space-time dependent fluctuations around the classical shape of the embedding. Namely, they are described by the fields $\theta (z, x^{\mu}) = \theta_{cl} (z) + \delta \theta (z,x^{\mu})$ and $\varphi (z,x^{\mu}) = \varphi_{cl} (z) + \delta \varphi (z,x^{\mu})$, where $x^{\mu}$ are the four-dimensional coordinates in (\ref{leadMetric}). To find the Lagrangian for these fields, one begins with the standard DBI action that describes the D7 branes, i.e.
\be
L = - \sqrt{- \det g_{8d}} 
\ee
with $g_{8d}$ being the eight-dimensional induced metric on the worldvolume, and substitutes (\ref{leadMetric}) with $\theta = \theta (z,x^{\mu})$, $\varphi = \varphi (z,x^{\mu})$. The resulting Lagrangian, up to an overall constant, is \cite{ASW2}:
\be\label{Lz2}
L \sim \sqrt{(1+z^2)\left[\theta_z{}^2+\sin^2\theta\,\varphi_z{}^2\left(1+\frac{c}{4}\theta_\m{}^2\right)\right]+\frac{\gamma\,z^2}{3}+\frac{c\,\gamma\,z^2}{12}\left[\sin^2 \theta \,\varphi_\mu{}^2+\theta_\mu{}^2\right]} \, ,
\ee
where we have introduced the notation $\gamma\equiv e^{4\,\rho_0}\sin^3(\alpha)<\!\!<1$. {}\footnote{Note that, in (2.14) of \cite{ASW2}, we have substituted with its background value the $\sin^2 \theta$ multiplier in front of the $\varphi_{\mu}^2$ term in the last bracket. This is because it leads to higher (than second) order in the fluctuations; see the discussion in Appendix A of \cite{ASW2}. Here we retain this multiplier for completeness.} The subscripts $z$ and $\mu$ on the functions $\theta$ and $\varphi$ denote derivatives with respect to $z$ and $x^{\mu}$, respectively. It is important to keep in mind that the theory we consider has a cutoff at a finite value of $z$, denoted by $z_\Lambda$, which corresponds to the physical cut-off $\rho_{\Lambda}$ mentioned above. Furthermore, though the range of $z$ is $-z_\Lambda\leq z\leq z_\Lambda$, it is sufficient to find solutions in the range $0\leq z\leq z_\Lambda$, due to the symmetric or anti-symmetric boundary conditions imposed at $z=0$.

One can easily verify \cite{ASW2, LA} that a simple classical solution of the Euler-Lagrange equations for $\theta$ and $\varphi$ is given by $\theta^{\rm cl}=\pi/2$ and
\be\label{varphi}
\varphi^{\rm cl}_z=\frac{\sqrt{\gamma}}{\sqrt{3}\,\sqrt{1+z^2}} \,.
\ee
To investigate the perturbative stability of this solution, one needs to study {\em small} $\delta \theta$ and $\delta \varphi$ fluctuations around this configuration. It is easy to see that the two kinds of fluctuations decouple to second order in the expansion of the Lagrangian.

Let us first consider the fluctuations of $\varphi$. To second order, the effective Lagrangian has the form:
\be\label{lquad}
L_{\rm quad}=\frac{\sqrt{3}\,z^2}{2\,\sqrt{\gamma}\,\sqrt{1+z^2}}\left\{\frac{\gamma\,c}{12}\,[\delta\varphi_\mu(z)]^2+[\delta\varphi'(z)]^2\right\}+\delta\varphi'(z) \, ,
\ee
where for convenience we have suppressed the dependence of $\delta\varphi$ on the space-time coordinates $x^\m$ and, also, we have introduced the notation $\varphi'= \pd_z \varphi$. For plane wave solutions with mass $m$, we can rewrite the Lagrangian as:
\be\label{lquad2}
L_{\rm quad}=\frac{\sqrt{3}\,z^2}{2\,\sqrt{\gamma}\,\sqrt{1+z^2}}\left\{M^2\,[\delta\varphi(z)]^2+[\delta\varphi'(z)]^2\right\}+\delta\varphi'(z) \, ,
\ee
where we have denoted $M^2 \equiv c\,\gamma\, m^2\,/\,12$ . The equation of motion, following from that Lagrangian, is:
\be\label{varphieq}
-\delta\varphi''(z)-\frac{2+z^2}{z\,(1+z^2)}\delta\varphi'(z)=M^2\delta\varphi(z) \, .
\ee

To be able to estimate analytically the spectrum at leading order, in \cite{ASW2} we rephrased the problem of solving (\ref{varphieq}) into a quantum mechanical problem for a Hamiltonian
\be
H_0 + \Delta H_0 \, ,
\ee
where
\be
H_0 = - \frac{d^2}{dz^2} - \frac{1}{z} \frac{d}{dz} \,\, ;
\ee
see Section 4 there. The key use of this was that the equation
\be \label{H0}
H_0 \,\chi (z) = M^2 \,\chi (z)
\ee
for the eigenfunctions $\chi$ and eigenvalues $M^2$ of $H_0$ can be easily solved analytically exactly. The solutions given in (4.14) of \cite{ASW2} are:
\be \label{chiSol}
\chi_n (z) = J_0 \left( \frac{r_n z}{z_{\Lambda}} \right) \, ,
\ee
where $J_0$ is the Bessel function. Having this result, one can use quantum mechanical perturbation theory to find the corrections that are due to the Hamiltonian $\Delta H_0$. The outcome is written down in \cite{ASW2}. That it does not contain any tachyon mode is due to the fact that the leading spectrum, which corresponds to (\ref{chiSol}), does not have any negative mass-squared modes. The correction due to $\Delta H_0$ is subleading in $\gamma$, compared to the eigenvalues $M^2$ of $H_0$, and therefore cannot change the positivity or negativity of the total answer for the mass-squared.

Note, however, that the solution (\ref{chiSol}) already assumes $M^2 > 0$. So to verify that there are no tachyonic modes in the above consideration, let us consider in more detail the case $M^2 < 0$. In such a case, the solution to (\ref{H0}) is the following combination of modified Bessel functions:
\be \label{modBess}
c_1 I_0 + c_2 K_0 \, ,
\ee
where $c_1$ and $c_2$ are arbitrary constants. However, the physical solutions in this model have to satisfy the conditions that they are regular\footnote{For more on this condition, see Appendix A.} and vanish at $z = z_{\Lambda}$.\footnote{This is precisely why the solution (\ref{chiSol}) does not contain the $Y_0$ Bessel function.} Now, it easy to realize that these conditions rule out any solution of the form (\ref{modBess}). More precisely, $K_0$ is singular at the origin $z=0$, while $I_0$ is monotonic and everywhere positive. This led us in \cite{ASW2} to the conclusion that there is no negative mass-squared mode in the spectrum. However, as we explain in a later section, the choice of coordinates used so far is not the most appropriate one. As a result, it misses the lowest fluctuation mode, which turns out to be a tachyon. Before turning to that, it will be useful and illuminating for the future to reconsider the numerical work of \cite{CLtV}, the first to indicate an instability in this model, via analytical means. 

To make the comparison to \cite{CLtV} more transparent, we will now study the Schr\"odinger form of the equation of motion. As pointed out in \cite{ASW2}, equation (\ref{varphieq}) can be transformed to the Schr\"odinger form
\be \label{Sch}
-\Phi'' (z) - \frac{6 + z^2}{4 (1+z^2)^2} \,\Phi (z) = M_{\varphi}^2 \Phi (z)
\ee
via the field redefinition
\be \label{FRed}
\delta \varphi (z) = \frac{(1+z^2)^{1/4}}{z} \Phi (z) \, .
\ee
In the next section we turn to studying equation (\ref{Sch}). We will find by analytical means the tachyonic mode of \cite{CLtV}. However, the latter will not translate to a physical mode, as it corresponds to a singular solution of (\ref{varphieq}), despite being a regular solution of (\ref{Sch}). The underlying reason for the physical inequivalence of the two formulations, i.e. equations (\ref{varphieq}) and (\ref{Sch}), is that the transformation (\ref{FRed}) is singular at $z=0$. The point $z=0$ is of crucial importance, as this is where one must impose the boundary conditions determining the spectrum in the Schr\"odinger picture.

Nevertheless, we will show in Section 5 that there is, after all, a tachyonic mode in this model. However, to see that one needs to use an improved choice of coordinates that separates properly the transverse and worldvolume directions with respect to the U-shaped probe D7-$\overline{{\rm D}7}$ embedding.

\section{Schr\"{o}dinger equation and stability}
\setcounter{equation}{0}

Let us rewrite equation (\ref{Sch}) as:
\be\label{eqpsi}
-\Phi''(z)+V(z)\,\Phi(z)=M^2\,\Phi(z) \, ,
\ee
where
\be\label{ourpotential}
V(z)=-\frac{6+z^2}{4(1+z^2)^2} \, .
\ee
This is in complete agreement with the Schr\"odinger equation and potential studied in \cite{CLtV}. The latter work used numerical methods to find the spectrum.

We would like to investigate (\ref{eqpsi}) with analytical means. In order to do that, we will use a set of theorems recalled in Appendix \ref{Ap}. The main point is that, to find whether or not there is a tachyonic mode, one only needs to look for zero mass solutions. If there is such a solution and it vanishes at some point $z_c > 0$, then there is a tachyonic mode for $z_{\Lambda} > z_c$. If there is no zero mass solution or the $M=0$ solution does not vanish anywhere, then the spectrum does not contain any tachyonic modes. The precise derivation of the relevant set of quantum mechanical theorems is given in Appendix \ref{Ap}. But the conceptual picture behind them is the following. The values of the mass levels, $m_n^2$, depend on the value of the cutoff $z_{\Lambda}$. More precisely, by decreasing $z_{\Lambda}$ one increases $m_n^2$ monotonically. Furthermore, at low enough $z_{\Lambda}$, all modes have positive mass-squareds $m_n^2$. Therefore, if by increasing sufficiently $z_{\Lambda}$, we can find a point $z_{\Lambda} = z_c$, such that for the lowest level $m_0^2 (z_c) = 0$, then when $z_{\Lambda} > z_c$ we will have $m_0^2 (z_{\Lambda}) < 0$ and thus a tachyon. Now, recall that a boundary condition for the physical solutions of (\ref{eqpsi}) is that they vanish at $z_{\Lambda}$. Hence, if there is a zero mass solution vanishing at a certain point $z_c$, then for $z_{\Lambda} > z_c$ the spectrum contains a tachyonic mode.

In view of the above, we now turn to studying the zero mass solutions of (\ref{eqpsi}). First, note that it has complete sets of solutions with both Neumann and Dirichlet boundary conditions at $z=0$. At the same time, every solution must satisfy the boundary condition $\delta\varphi(z_\Lambda)=0=\Phi(z_\Lambda)$. For $M=0$ the general solution of this equation is
\be\label{zeromass}
\Phi_0(z)=c_D \,\frac{z}{(1+z^2)^{1/4}}+c_N\,\frac{z\,\sinh^{-1}(z)-\sqrt{1+z^2}}{(1+z^2)^{1/4}} \, ,
\ee
where $c_D$ and $c_N$ are integration constants. Clearly, for Neumann boundary condition at $z=0$ one has $c_D=0$, while for Dirichlet boundary condition $c_N=0$. While the latter solution does not vanish at any $z>0$, the solution with Neumann boundary condition has a zero at $z=z_c\simeq 1.5$. Thus, there is no zero mass solution with Dirichlet boundary conditions, but there is a zero mass state for the Neumann case if we choose $z_\Lambda=z_c$. As remarked earlier though, the physical choice is $z_\Lambda>\!\!>1$. Since $z_{\Lambda} > z_c$ then according to the theorems, proven in  Appendix \ref{Ap}, there must exist a tachyonic solution, i.e. with $M^2<0$. Thus the model seems to be perturbatively unstable. This is in complete agreement with the numerical result of \cite{CLtV} that instability sets in, when the cutoff is taken to be larger than $z_c=1.5$.

However, the mode with $M^2 < 0$ is singular when transformed back to a solution of equation (\ref{varphieq}). Indeed, applying (\ref{FRed}) to (\ref{zeromass}), we find that it acquires the form:
\be\label{massless}
\delta\varphi_0(z)=c_D+c_N\,\left[\sinh^{-1}(z)-\frac{\sqrt{1+z^2}}{z}\right].
\ee
This solution diverges at $z=0$ and the key reason for that is the singularity of the transformation (\ref{FRed}) at the origin. As a result, the solution (\ref{massless}) would be ruled out by the regularity condition\footnote{The regularity condition is well-known in the literature. Nevertheless, for more completeness, in Appendix A we give some arguments as to why singular solutions are inadmissible.} that needs to be imposed, just as was the case in the language of the previous section. However, as already alluded to above, the coordinates used so far are not the best suited for properly capturing all the information about the scalar spectrum. We will discuss a better choice in Section 5 and show that there is a tachyonic mode after all. 

As a final comment, it is easy to show by series expansion that the massive solutions of (\ref{varphieq}) with Neumann boundary condition in the Schr{\"o}dinger form also have a pole at $z=0$, just like the massless one.

\section{The $\delta \theta$ spectrum revisited}
\setcounter{equation}{0}

In this section, we will reconsider the fluctuations of $\theta$ with the new method developed here. As we have shown in \cite{ASW2}, the equation of motion for $\delta\theta$, that follows from the Lagrangian (\ref{Lz2}), can be converted to Schr\"odinger form via the field redefinition
\be \label{TTr}
\delta \theta (z) = \frac{\Theta (z)}{(1+z^2)^{1/4}} \, .
\ee
The result is:
\be\label{thetaeq}
-\Theta''(z)-\left( \frac{z^2-2}{4\,(1+z^2)^2} + \frac{\gamma}{3\,(1+z^2)} \right) \Theta (z) =0 \, ,
\ee
where we have set the mass to zero in view of the method of the previous section. Note that  transformation (\ref{TTr}) is regular everywhere, unlike the $\delta \varphi$ transformation (\ref{FRed}). Thus the original equation of motion in the $\theta$ case is physically completely equivalent to the Schr\"odinger form one. 

\begin{figure}[htbp]
\begin{center}
\includegraphics[width=4in]{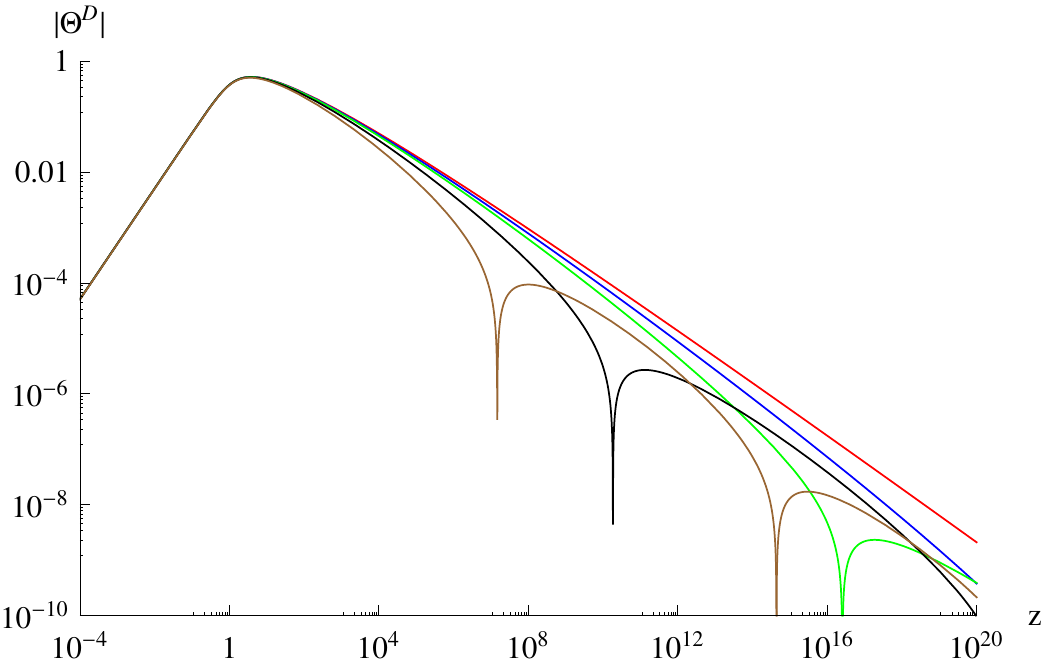}
\caption{Plot of the absolute value of $\Theta^D$ as a function of $z$ at the following choices of $\gamma$: 0.1 (brown), 0.05 (black), 0.02 (green), 0.01 (blue), and 0.001 (red).  The spikes in the plot correspond to zeros of $\Theta^D$.}
\label{dirichlet1}
\end{center}
\end{figure}

\begin{figure}[htbp]
\begin{center}
\includegraphics[width=4in]{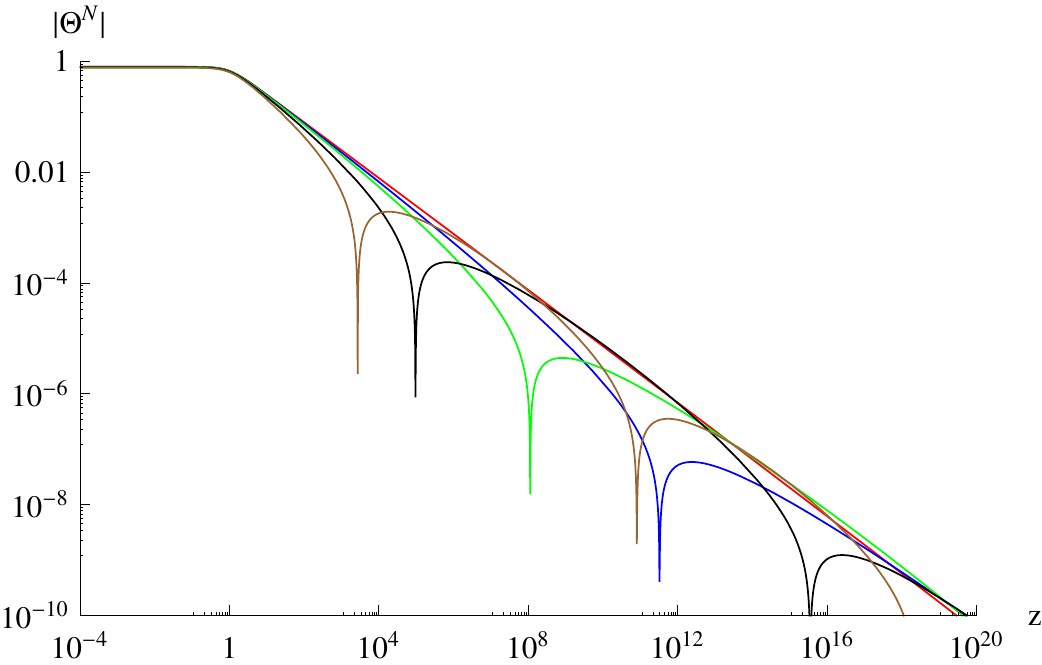}
\caption{Plot of the absolute value of $\Theta^N$ as a function of $z$ at the following choices of $\gamma$: 0.1 (brown), 0.05 (black), 0.02 (green), 0.01 (blue), and 0.001 (red).  The spikes in the plot correspond to zeros of $\Theta^N$.}
\label{neumann1}
\end{center}
\end{figure}

Now, the general solution of (\ref{thetaeq}) is:
\be
\Theta(z)=c_D\,\Theta^D(z)+c_N\,\Theta^N(z) \, ,
\ee
where the odd Dirichlet solution and the even Neumann solution are combinations of Legendre functions, namely:
\bea \label{thetasolution}
\Theta^D(z)&=&i\,P_{i\,\nu-1/2}^{1/2}(i\,z)+\frac{2}{\pi}\,\coth\left(\frac{\pi\,\nu}{2}\right)\, Q_{i\,\nu-1/2}^{1/2}(i\,z)\nn\\
\Theta^N(z)&=&i\,P_{i\,\nu-1/2}^{1/2}(i\,z)+\frac{2}{\pi}\,\tanh\left(\frac{\pi\,\nu}{2}\right)\, Q_{i\,\nu-1/2}^{1/2}(i\,z)
\eea
with $\nu=\sqrt{\gamma\,/\,3}$.
Despite the appearance of the imaginary unit in (\ref{thetasolution}), both $\Theta^D$ and $\Theta^N$ are real. In Figs. \ref{dirichlet1} and \ref{neumann1} we have plotted the absolute values of $\Theta^D$ and $\Theta^N$ in (\ref{thetasolution}) for a series of choices of $\gamma$. The spikes of the absolute values correspond to zeros of the functions. We should note that, for illustrative purposes, we have used values of $\gamma$ that are (significantly) larger than the phenomenologically acceptable range (see \cite{ASW1}). This is because for smaller values of $\gamma$ the zeros of  $\Theta^D$ and $\Theta^N$ occur at extremely large $z$, which makes plotting very difficult.

Nevertheless, the asymptotic form (at large $z$) of (\ref{thetasolution}) makes the location of the zeros of $\Theta^D$ and $\Theta^N$ rather clear. Namely, one can show that at large $z$:
\bea\label{thetaass}
\Theta^D(z)&\sim&\frac{1}{\sqrt{z}}\sin\left[\sqrt{\frac{\gamma}{3}}\log(z)\right],\nn\\
\Theta^N(z)&\sim&\frac{1}{\sqrt{z}}\cos\left[\sqrt{\frac{\gamma}{3}}\log(z)\right].
\eea
The location of the zeros, in the limit of small $\gamma$, can then be read off from (\ref{thetaass}). In particular, the smallest zeros correspond to the following values of $z$:
\bea\label{thetainstability}
z_c^D&=&e^{\pi\,\sqrt{3\,/\,\gamma}} \, ,\nn\\
z_c^N&=&e^{\pi\,/\,2\,\sqrt{3\,/\,\gamma}} \, .
\eea
Then, according to the theorems in Appendix B, an instability would occur if one takes the cutoff to be $z_{\Lambda}^{D,N} > z_c^{D,N}$. Note, however, that the lower limits on the cutoffs, i.e. (\ref{thetainstability}), comfortably exceed the values of $z_\Lambda$ required for phenomenological reasons. More precisely, the latter means the requirement that the $S$-parameter in this model is consistent with observation. Hence, our phenomenological fits in \cite{ASW1} are in the region of stability.

Note that the presence of an instability in the $\theta$ sector, for a large enough cutoff, was   found in \cite{CLtV} with an entirely different method; see the discussion of Model 1 in their  Appendix. Our estimate (\ref{thetainstability}) for the critical cut-off, above which it sets in, matches perfectly with the one given in their equation (44).

Finally, let us comment on why the theoretical instability in the $\delta\theta$ sector was not discovered in \cite{ASW2}. The reason is that $z_c^N$ and $z_c^D$ are non-perturbative functions of $\gamma$ and, in particular, $z_c^{N,D}\to\infty$ when $\gamma\to0$. In \cite{ASW2}, we treated the correction term in (\ref{thetaeq}), namely the one proportional to $\gamma$, as a small perturbation. Therefore, results non-perturbative in $\gamma$ were beyond reach for the methods used there. 

\section{Cartesian coordinates and instability}
\setcounter{equation}{0}

In the previous sections we described the U-shaped embedding of the probe branes as a function $\varphi (z)$, where the worldvolume radial coordinate $z$ is related to the spacetime one $\rho$ via $z = \pm \sqrt{e^{4 (\rho - \rho_0)} - 1}$; see \cite{ASW2}. However, the 10d spacetime coordinate $\varphi$ is not always transverse to the U-shaped embedding. Therefore, it may not capture the full information about the scalar field spectrum, which arises from fluctuations of the transverse coordinates. Indeed, close to the tip, which is given by $\rho = \rho_0$, the radial coordinate $\rho$ is a better approximation to a transverse coordinate, whereas far away from the tip the angular coordinate $\varphi$ plays such a role. Hence, to find an appropriate choice, such that one coordinate remains transverse and the other one remains longitudinal  everywhere along the embedding, we need to perform a change of variables that mixes $\varphi$ and $\rho$.

It turns out that the appropriate new coordinates can be defined by:
\bea \label{yzhat}
y^2 &=& e^{4\rho} \left( 1 - \tanh^2 \left( \frac{\sqrt{3} \,\varphi }{\sqrt{\beta} \,e^{2 \rho_0}} \right) \right) \, , \nn \\
\hat{z}^2 &=& e^{4\rho} \,\tanh^2 \left( \frac{\sqrt{3} \,\varphi}{\sqrt{\beta} \,e^{2 \rho_0}} \right) \, .
\eea
The reason is the following. The classical embedding solution, as a relation between $\varphi$ and $\rho$, is given by \cite{LA}:
\be \label{clsol}
\tanh \!\left( \frac{\sqrt{3} \,\varphi (\rho)}{\sqrt{\beta} \,e^{2\rho_0}} \right) = \pm \sqrt{1- e^{4(\rho_0 - \rho)}} \,\,\, .
\ee
Rewriting this as:
\be \label{clsol2}
e^{4\rho_0} = e^{4\rho} \left( 1 - \tanh^2 \left( \frac{\sqrt{3} \,\varphi (\rho)}{\sqrt{\beta} \,e^{2 \rho_0}} \right) \right) \, ,
\ee
one can see that the coordinate $y$, defined in (\ref{yzhat}), is constant everywhere along the embedding. Namely, it satisfies
\be\label{classicaly}
y = e^{2\rho_0}
\ee
for every point belonging to the classical solution. Therefore, the canonical transverse fluctuations are the fluctuations of $y$. On the other hand, the coordinate $\hat{z}$ in (\ref{yzhat}) describes the position of a point on the embedding. So we can use $\hat{z}$ directly as the worldvolume radial coordinate. In that regard, note that it obviously runs from $-\infty$ to $+\infty$ and, furthermore, when $\rho = \rho_0$ we have $\varphi = 0$, which implies $\hat{z}=0$. Finally, let us also point out
that (\ref{yzhat}) implies the "Cartesian" relation:
\be
y^2 + \hat{z}^2 = e^{4\rho} \, ,
\ee
which is rather similar to the situation in both \cite{KS} and \cite{SS}, after they introduced new coordinates to study the scalar spectrum. 

Now we turn to considering the DBI action for the probe D7-$\overline{{\rm D}7}$ embedding. To investigate the scalar spectrum, we promote the transverse coordinate $y$ to a field on the worldvolume $y (\hat{z},x^{\mu}) = e^{2 \rho_0} + \delta y (\hat{z} , x^{\mu})$, where $\delta y$ is the fluctuation around the classical solution $y_{cl} = e^{2 \rho_0}$. Expanding the DBI action to second order in the fluctuations, we obtain the following quadratic Lagrangian:\footnote{For consistency, one can check first that $y = e^{2 \rho_0}$ is indeed a classical solution. For that purpose, we note that the DBI action for the $x^{\mu}$-independent classical configuration $y (\hat{z})$ is $L=\frac{1}{y}\sqrt{e^{4\,\rho_0}[y-\hat z\,y']^2+y^2\,[\hat z+y\,y']^2}$. It is easy to verify that its equation of motion is satisfied by $y = e^{2 \rho_0}$.}
\be\label{Lquad}
L_{\rm quad}=- \frac{4\,\hat z\,\delta y\,\delta y'}{\sqrt{e^{4\,\rho_0}+\hat z^2}}-\sqrt{e^{4\,\rho_0}+\hat z^2}\,\left[\delta y'{}^2+\frac{c\,\beta}{12}\delta y_\mu{}^2\right],
\ee
where $\delta y_\mu{}^2$ is the derivative with respect to the spacetime coordinates $x^{\mu}$.
The equation of motion for $m=0$ fluctuations is then 
\be
(e^{4\,\rho_0}+\hat z^2)^2\delta y''+(e^{4\,\rho_0}+\hat z^2)\,\hat z\,\delta y'+2\,e^{4\,\rho_0}\,\delta y=0 \, .
\ee
The general solution of this equation is
\be
\delta y= c_D\,\frac{\hat z}{\sqrt{e^{4\,\rho_0}+\hat z^2}}+c_N\, \frac{\hat z \,\log[e^{-2\,\rho_0}\,(\hat z+\sqrt{e^{4\,\rho_0}+\hat z^2})]-\sqrt{e^{4\,\rho_0}+\hat z^2}}{\sqrt{{e^{4\,\rho_0}+\hat z^2}}} \, .
\ee
The solution with Neumann boundary condition at $\hat z=0$ has a zero at finite $\hat z=\hat z_c$, implying, according to the set of theorems in Appendix B, that the classical solution is unstable if the cutoff $\hat z_\Lambda >\hat z_c\simeq 1.5\times e^{2\,\rho_0}$.

One can make the instability more explicit by diagonalizing the quadratic effective Hamiltonian for fluctuations. Introducing $\delta Y=\delta  y\,\left( e^{4\,\rho_0}+\hat z^2 \right)$ in (\ref{Lquad}) we obtain for the quadratic Hamiltonian
\be\label{Hquad}
H_{\rm quad}\sim \int\frac{d\hat z\,d^4x}{(e^{4\,\rho_0}+\hat z^2)^{3/2}}\left[\delta Y'{}^2 -\frac{c\,\beta}{12}\sum_{\mu=0}^3\left(\partial_\mu \delta Y\right)^2-\frac{4\,\hat z^2}{(e^{4\,\rho_0}+\hat z^2)^2}\,\delta Y^2\right].
\ee
The negative signs of the last two terms in (\ref{Hquad}) are in contrast to the similar expression for the Sakai-Sugimoto model \cite{SS}, in which all terms have positive signs. This difference has a crucial implication. Namely, the Hamiltonian in our case must have a negative eigenvalue, which signifies instability. To realize this, note that the Hamitonian (\ref{Hquad}) takes a negative value at the trial function $\delta Y = const$.

Finally, let us discuss the broader issue of how much the entire mass spectrum, obtained from working in the Cartesian coordinate system introduced in this section, can differ from the spectrum obtained by using the old coordinate system. To answer this question, notice the following. Upon substituting in (\ref{Lquad}) the ansatz $\delta y (\hat{z}) = (1+ \zeta^2)^{-1/4} \,\delta u (\zeta)$, where $\zeta = \hat{z} e^{-2\rho_0}$, one can verify that the resulting equation of motion becomes identical to (\ref{eqpsi}), with $\Psi$ replaced by $\delta u$ and $z$ by $\zeta$. Therefore, the two spectra are identical, modulo some modes that may be present in one of them but missing in the other due to different boundary and/or normalizability conditions in the two coordinate systems. We have already seen that the tachyon is such a mode, present in the Cartesian coordinate system but missing in the old one. It is worth noting that the old coordinate system may, in principle, be missing other modes as well, since it does not split properly between transverse and longitudinal directions to the probe brane's worldvolume.

\section{Conclusions}

We reconsidered the problem of stability of the D7-$\overline{{\rm D}7}$ embeddings studied in \cite{ASW2,CLtV}. We developed a new analytical method, that allowed us to reproduce the numerical results of \cite{CLtV}. The method relies on the connection between tachyons and zero mass modes in a Schr\"{o}dinger equation with a bounded potential and a finite range for the variable. Furthermore, we pointed out that, although the equations of motion studied in \cite{ASW2} and in \cite{CLtV} are mathematically equivalent, they do not capture the same physical information in the $\varphi$ sector. The reason is that the transformation, relating them to each other, is singular at the point $z=0$, where a crucial boundary condition needs to be imposed. The appropriate form of the $\delta \varphi$ equation of motion does not contain a tachyonic mode. 

Nevertheless, the model is perturbatively unstable. The information about the physical tachyon is not visible in the description provided by the field $\varphi$, because the spacetime direction parametrized by the coordinate $\varphi$ is not transverse everywhere along the D7-$\overline{{\rm D}7}$ embedding. We introduced a suitable change of variables, leading to new coordinates that are separating properly the transverse and longitudinal directions. Using these, we showed that the scalar spectrum does contain a tachyonic mode. Up to the Jacobian of the relevant coordinate transformation, that mode agrees with the tachyon of \cite{CLtV}. It would be very interesting to explore stabilizing the model by turning on worldvolume flux as in \cite{MW,BJLL}. We hope to come back to this question soon.

It is also interesting to note that, with our new analytical method, we confirmed the presence of a $\theta$ sector instability, in agreement with \cite{CLtV}. More precisely, we showed that the spectrum of $\delta \theta$ fluctuations contains a tachyonic mode, when the cutoff of the model exceeds a certain value. This value is extremely large, being given by an exponential of an inverse power of the small parameter $\gamma$. Thus, it is well beyond what is allowed by the requirement for a phenomenologically acceptable value of the S-parameter \cite{ASW1}, when this model is viewed as realizing dynamical electroweak symmetry breaking.\footnote{A clear caveat is that one needs to investigate first whether the presence of worldvolume flux modifies the phenomenological constraint or not.} Nevertheless, this is an instability that exists in the theoretical model, studied here, of a flavored strongly coupled gauge theory with two dynamical scales, which are separated by a nearly conformal region. It is conceivable that this gauge theory might have other applications in the future\footnote{For example, an intriguing new application of the walking background of \cite{piai} is within the context of the recently proposed model of slow-walking Cosmological Inflation \cite{EHNT}.}, in which the new instability could play a role.

\section*{Acknowledgements}

We are very grateful to T. ter Veldhuis for useful correspondence and to P. Argyres and G. Semenoff for fruitful discussions. We also thank M. Piai for interesting conversations. Research at Perimeter Institute is supported by the Government of Canada through Industry Canada and by the Province of Ontario through the Ministry of Research \& Innovation. L.A. is also supported in part by funding from NSERC. The research of R.W. is supported by DOE grant FG02-84-ER40153.

\appendix

\section{On singular solutions for $\delta\varphi$}
\setcounter{equation}{0}

In this Appendix we review a couple of arguments as to why one needs to impose the regularity condition for solutions of the equations of motion in our context. 
It should be kept in mind that the appendix relies entirely on the coordinate system used by \cite{ASW2,CLtV}, in which the fluctuating field does not have a transverse component {\it everywhere} along the D7-$\overline{{\rm D}7}$ brane embedding. Therefore, one is not guaranteed to capture the full spectrum. In Section 5 of the main text, we will address this issue by introducing new coordinates, in which the fluctuating field is transverse everywhere along the embedding. In the new coordinates, we will find a tachyonic mode given, up to the relevant Jacobian factor, by the same expression as the one discussed here and in \cite{CLtV}.  

As we have seen in Section 3, a singular solution of (\ref{varphieq}) arises from solving (\ref{eqpsi}) with Neumann boundary condition at $z=0$. The first argument against accepting such a solution as part of the physical spectrum is the requirement for a finite action for any perturbative mode.
To see how the latter is violated by the singular solutions, let us consider the kinetic term for $\delta \varphi$ fluctuations in the action:
\be\label{LH}
S=\int dz\, \frac{\sqrt{3}\,z^2}{2\,\sqrt{\gamma}\,\sqrt{1+z^2}}\,\left[\delta\varphi'(z)\right]^2+ ... \,\,\, .
\ee
As we noted earlier, all solutions for $\delta\varphi$ with Neumann boundary conditions in Schr\"odinger form have the same behavior at $z=0$ as the massless solution of (\ref{massless}). In other words, near $z=0$ one has $\delta\varphi'(z)\sim z^{-2}$. Now, upon substituting back such a solution in (\ref{LH}), one clearly obtains a divergent answer. Thus, requiring finite action rules out those solutions.\footnote{One should keep in mind though, that the consideration here refers to the coordinate system used in \cite{ASW2,CLtV}. As already mentioned above, in Section 5 we will introduce a more suitable coordinate system that will lead to a different conclusion regarding the tachyon mode.}

Another argument for requiring regular solutions is the following. The quadratic action, on which the analysis of perturbative (in)stabilities relies, should not break down as the leading approximation to the full action. To see that in our case there is such a break down, let us consider in some detail the power series expansion of the quadratic action in a neighborhood of the point $z=0$.
First, note that the terms containing $\delta\varphi_\mu$ in (\ref{Lz2}) have a pre-factor $z^2$. Therefore, when calculating the expansion coefficients of $\delta\varphi'^n$ near $z=0$, we may neglect $\delta\varphi_\mu$ altogether. For simplicity, we can also neglect contributions from $\delta\theta$, which leaves us with the following approximate action:
\be\label{Lz0}
L_z^0\sim\sqrt{ \left[\delta\varphi'(z)+\varphi_{\rm cl}'(z)\right]^2+\frac{\gamma\,z^2}{3}}
\ee
valid in a neighborhood of $z=0$. Now, expanding $L_z^0$ in $\delta\varphi'(z)$, we arrive at the following schematic expression for the action near $z=0$:
\be
  L \simeq  \frac{z^2}{2\,\varphi_{\rm cl}'(z)} \,\sum_n \left[-\frac{\delta\phi'(z)}{\varphi_{\rm cl}'(z)}\right]^n\simeq \frac{z^2}{2 \,\varphi_{\rm cl}'(z)} \,\sum_n [-\delta\phi'(z)]^n\,\left(\frac{3}{\gamma}\right)^{n/2}.
  \ee
This power series is convergent only if $\delta\varphi'(z) <\varphi_{\rm cl}'(z)\simeq \sqrt{\gamma\,/\,3}$\,, where the last relation follows from (\ref{varphi}). The convergence  condition is clearly violated for all solutions that are singular at $z=0$. In particular, this means that it is violated for all solutions of the Schr\"odinger form equation, which satisfy a Neumann boundary condition at $z=0$. As we have already seen, the tachyonic mode in the $\delta \varphi$ spectrum belongs exactly to this class of solutions. In other words, it is outside the range of validity of the action, whose equations of motion it solves. Thus, it is not clear whether it is a legitimate solution of the physical problem at hand.

Note that, although in our case the first argument we gave above is, clearly, stronger than the second one, the second argument is worth mentioning for the following completeness reason. In other examples, which could have solutions with logarithmic singularities, the requirement for finite action alone would not be enough to rule out singular solutions.

\section{Schr\"odinger equation: a set of theorems} \label{Ap}
\setcounter{equation}{0}

In this appendix we review several quantum mechanical theorems, that are of crucial importance for the method used in Sections 3 and 4. Namely, those theorems establish the connection between zero mass modes and negative mass solutions of the Schr\"odinger equation under certain conditions.

Let us consider the Schr\"odinger equation
\be\label{eq}
-\psi''(z)+V(z)\,\psi(z)=E\,\psi(z)
\ee
with a potential $V(z)$ that is a real analytic function of $z$ for all $z\geq 0$. In fact, it will be enough to require the considerably weaker condition of piecewise analyticity and continuity of $V(z)$, together with the constraint $|r\,V(r)|<\infty$ at $z=0$. We will take the variable $z$ to vary within the interval $0\leq z\leq z_\Lambda$. Also, we will impose the boundary condition that at the cutoff $z_\Lambda$ all eigenfunctions vanish, i.e. $\psi_n(z_\Lambda)=0$. Upon imposing appropriate boundary conditions at $z=0$ (the form of which will be discussed below), equation (\ref{eq}) has a discrete spectrum with infinitely many eigenvalues. We will investigate the dependence of those eigenvalues on $z_\Lambda$. In particular, we will address the question of when there are eigenfunctions with negative eigenvalues $E<0$, whose presence would indicate an instability.

In what follows, we will present a series of theorems, which allow us to reach the conclusion that the instability for a cutoff $z_\Lambda$ is related to the existence of a zero mass solution for a smaller cutoff $z_\lambda<z_\Lambda$. We do not claim that all of these theorems are original, but we list them for completeness.

 {\bf Theorem I:} Complete systems of eigenfunctions of (\ref{eq}) must satisfy either the Dirichlet $\psi(0)=0$ or the Neumann $\psi'(0)=0$ boundary conditions.

  Indeed, if two eigenfunctions $\psi_a$ and $\psi_b$ with eigenvalues $E_a\not=E_b$ are to be orthogonal, then
  \bea\label{relation}
\left[\psi_a\,,\,\psi_b\right]&\equiv&\lim_{z\to0}\,\left[ \psi_a(z)\,\psi_b'(z)-\psi_b(z)\,\psi_a'(z)\right]\nn\\&=&(E_b-E_a)\int_0^{z_\Lambda}\,\psi_a(z)\,\psi_b(z)\,dz=0 \, .
\eea
Hence, all the eigenfunctions can be divided into two complete orthogonal sets satisfying $\psi(0)=0$ and $\psi'(0)=0$ boundary conditions, respectively. And these are the only admissible boundary conditions at $z=0$.

 {\bf Theorem II:} Consider two potentials $V_0 (z)$ and $V_1 (z)$. Suppose that $V_1(z)\geq V_0(z)$ for every $z$ and that, at least in one finite interval, we have $V_1(z)>V_0(z)$. Let the ground state energy of equation
 \be
 -\psi''(z)+V_1(z)\,\psi(z)=E\,\psi(z)
 \ee
 be $E_1$ and the ground state energy of equation
 \be
 -\psi''(z)+V_0(z)\,\psi(z)=E\,\psi(z)
 \ee
be $E_0$. Then, the inequality $E_1>E_0$ holds.

Indeed, this statement is true for the following reason. If we denote the ground state eigenfunctions by $\psi_1$ and $\psi_0$ respectively, then we have:
\be
 E_1=\frac{\int_0^{z_\Lambda}dz\, \left\{\left[\psi_1'(z)\right]^2+V_1(z)\,[\psi_1(z)]^2\right\}}{\int_0^{z_\Lambda}dz\, V_1(z)\,[\psi_1(z)]^2} \, .
 \ee
Let us consider the expression
\bea
E_2&=&\frac{\int_0^{z_\Lambda}dz\, \left\{\left[\psi_1'(z)\right]^2+V_1(z)\,[\psi_1(z)]^2\right\}}{\int_0^{z_\Lambda}dz\, [\psi_1(z)]^2}-\frac{\int_0^{z_\Lambda}dz\, [V_1(z)-V_0(z)]\,[\psi_1(z)]^2}{\int_0^{z_\Lambda}dz\, [\psi_1(z)]^2}\nn\\&=&\frac{\int_0^{z_\Lambda}dz\, \left\{\left[\psi_1'(z)\right]^2+V_0(z)\,[\psi_1(z)]^2\right\}}{\int_0^{z_\Lambda}dz\, [\psi_1(z)]^2}<E_1 \, .
\eea
Hence $E_0 <E_2 <E_1$, because $\psi_1(z)$ must be a mixture of the ground state and excited states of the equation with potential $V_0$. Therefore, the ground state eigenvalue is lowered by decreasing the potential.

{\bf Theorem III:} The ground state energy for a potential, bounded from below, is positive at a sufficiently small value of the cutoff $z_\Lambda$.

This theorem can be proven as follows. First, let us denote by $-N$ the lower bound of the potential $V$. Now, according to Theorem II, the ground state energy for $V$ is larger than that for the potential 
\be
\tilde V =\begin{cases} -N & \text{\,\, for \, $0\leq z\leq z_\Lambda$ \,,}\\
 \infty & \text{\,\, for \, $z>z_\Lambda$ \,.}\end{cases}
\ee
However, the ground state energy for $\tilde V$ is:
\be
\tilde E =\begin{cases} \frac{\pi^2}{z_\Lambda{}^2}-N & \text{\,\, for Dirichlet boundary condition at $z=0$ \,,}\\
 \frac{\pi^2}{4\,z_\Lambda{}^2}-N & \text{\,\, for Neumann boundary condition at $z=0$ \,.}\end{cases}
 \ee
Clearly, in both cases the ground state energy is larger than 0 at sufficiently small $z_\Lambda$.

Of course, this theorem is essentially a consequence of the Heisenberg uncertainty relation in free space. Note also that the same statement can be shown to be valid for a Coulomb like potential. However, the statement of this theorem is not, in general, valid for more singular potentials, such as $V(z)\sim z^{-2}$.

{\bf Theorem IV:} The ground state energy is a monotonically decreasing function of $z_\Lambda$.  

{\bf Proof:} To prove this theorem, we will consider two cutoffs $z_1$ and $z_2$, such that $z_2 > z_1$. Let us denote by $\psi^0(z,z_1)$ the ground state wave function for the cutoff $z_1$. As before, $\psi^0(z,z_1)$ is a real analytic function of $z$ for all $0<z<z_1$. Let us denote the corresponding ground state energy by $E_{z_1}^0$. Then we have: 
\be
\frac{\int_0^{z_1}\left\{\left[\psi_{,z}^0(z,z_1)\right]^2+V(z)\left[\psi^0(z,z_1)\right]^2\right\}
 \,dz}{\int_0^{z_1}\left[\psi^0(z,z_1)\right]^2\,
 dz}=E_{z_1}^0 \, .
\ee
Now, for an arbitrary cutoff $z_2>z_1$, we introduce a function $\tilde\psi(z,z_2)$, such that:
\be
\tilde\psi(z,z_2)=\begin{cases}\psi^0(z,z_1) & \text{ \,\, if $ z\leq z_1$ \,,}\\
0 & \text{ \,\, if $z_1\leq z\leq z_2$ \,.}
\end{cases}
\ee
Then, we also have:
\be
\tilde I=\frac{\int_0^{z_2}\left\{\left[\tilde\psi_{,z}(z,z_2)\right]^2+V(z)\left[\tilde\psi(z,z_2)\right]^2\right\}
 \,dz}{\int_0^{z_2}\left[\tilde\psi(z,z_2)\right]^2\,
 dz}=E_{z_1}^0
\ee

Since the wave function $\tilde\psi(z,z_2)$ vanishes at $z=z_2$, it can be expanded in a complete series of eigenfunctions, $\psi^k(z,z_2)$, defined on the interval $0\leq z\leq z_2$:
\be\label{expansion}
\tilde\psi(z,z_2)=\sum_k c_k \psi^k(z,z_2) \, .
\ee
Let us also denote by $E_{z_2}^k$ the eigenvlues of $\psi^k (z,z_2)$. Now,  $\tilde\psi(z,z_2)$ cannot be equal to the ground state wave function $\psi^0(z,z_2)$ for cutoff $z_2$. The reason is that the latter is an analytic function at $z=z_1$, while $\tilde\psi(z,z_2)$ has a discontinuous derivative at $z=z_1$. Thus,
 not all of the coefficients $c^k$, with $k>0$, in (\ref{expansion}) vanish. Hence:
\be
E_{z_1}^0=\tilde I= \frac{\sum_k |c_k|^2 E_{z_2}^k}{\sum_k |c_k|^2 }>E_{z_2}^0 \, .
\ee
So, indeed, the ground state energy is a monotonically decreasing function of the cutoff $z_\Lambda$.

{\bf Theorem V:} There is at least one tachyon in the spectrum for all cutoffs greater than a certain value $z_c$, i.e. for $z_\Lambda>z_c$, if and only if there is a zero mass solution for $z_\Lambda=z_c$, which solution satisfies either $\psi = 0$ or $\psi' = 0$ boundary conditions at $z=0$.

Indeed, according to Theorem III, at sufficiently low values of $z_\Lambda$ the ground state energy $E^0_{z_{\Lambda}}$ is positive. In addition, according to Theorem IV, the energy $E^0_{z_{\Lambda}}$ decreases as the cutoff $z_\Lambda$ is increased. Hence, there are two possibilities upon increasing $z_\Lambda$. The first is that the ground state energy remains always positive, i.e $\lim_{z_{\Lambda}\to\infty}E^0_{z_\Lambda}>0$. In this case, for all $z_\Lambda>0$ all eigenvalues $M^2=E^k_{z_{\Lambda}}>0$. Hence, there are no tachyons and the model is stable. The second possibility is that there is a critical value $z_c$, such that when $z_\Lambda=z_c$ we have $E_{z_{\Lambda}}^0=0$. Then, according to Theorem IV, whenever $z_\Lambda>z_c$ there must be at least one solution with $M^2=E<0$.

\end{document}